\documentclass[%
 reprint,
 amsmath,amssymb,
 aps,
pre,
]{revtex4-2}

\usepackage{graphicx}
\usepackage{dcolumn}
\usepackage{bm}
\usepackage{hyperref}

\usepackage[mathlines]{lineno}

\newcommand{\Eq}[1]{Eq.~(#1)}

\newcommand{\rmd}{\mathrm{d}}

\usepackage{xcolor}

\usepackage[normalem]{ulem}   

\definecolor{brickred}{rgb}{0.8, 0.25, 0.33}

\begin{document}


\title{Cost Functions in Economic Complexity}

\author{Alessandro Bellina}
\email{alessandro.bellina.1997@gmail.com}
\affiliation{%
 Centro Ricerche Enrico Fermi, Piazza del Viminale, 1, I-00184 Rome, Italy
}%
\affiliation{``Sony Computer Science Laboratories - Rome, Joint Initiative CREF-SONY, Centro Ricerche Enrico Fermi, Via Panisperna 89/A, 00184, Rome, Italy''}
\affiliation{Sapienza University of Rome, Physics Dept., P.le A. Moro, 5, I-00185 Rome, Italy}
\author{Paolo Butt\`a}%
\affiliation{%
 Sapienza University of Rome, Mathematics Dept., P.le A. Moro, 5, I-00185 Rome, Italy
}%


\author{Vito D.~P.~Servedio}
\affiliation{
 Complexity Science Hub, Metternichgasse 8, 1030 Vienna, Austria
}%


\begin{abstract}
\noindent 
Economic complexity algorithms aim to uncover the hidden capabilities that drive economic systems. Here, we present a fundamental reinterpretation of two of these algorithms, the Economic Complexity Index (ECI) and the Economic Fitness and Complexity (EFC), by reformulating them as optimization problems that minimize specific cost functions. 
We show that ECI computation is equivalent to finding eigenvectors of the network's transition matrix by minimizing the quadratic form associated with the network's Laplacian. 
For EFC, we derive a novel cost function that exploits the algorithm's intrinsic logarithmic structure and clarifies the role of the regularization parameter in its non-homogeneous version. Additionally, we establish the existence and uniqueness of its solution, providing theoretical foundations for its application.
This optimization-based reformulation bridges economic complexity and established frameworks in spectral theory, network science, and optimization. The theoretical insights translate into practical computational advantages: we introduce a conservative, gradient-based update rule that substantially accelerates algorithmic convergence, with potential implications for a broader class of algorithms, including the Sinkhorn-Knopp method. 
Finally, we apply the energetic framework to a real-world trade network, demonstrating how link-wise energy provides a direct way to identify structurally relevant and vulnerable regions of the export matrix, thus complementing and enriching standard economic complexity analyses. Beyond advancing our theoretical understanding of economic complexity indicators, this work opens new pathways for algorithmic improvements and extends applicability to general network structures beyond traditional bipartite economic networks.
\end{abstract}

\maketitle

\section{Introduction}
\noindent
    Understanding the productive capabilities of countries and the structure of their trade networks is fundamental to explaining patterns of economic growth and development.
    Economic complexity measures are designed to characterize the productive capabilities of countries and the structure of international trade~\cite{tacchella2012new, hidalgo2021economic, bahrami2023economic}. A first attempt to formalize this idea was made by Hidalgo and Hausmann through the Economic Complexity Index (ECI)~\cite{hidalgo2009building, hidalgo2007product}, based on a linear spectral method applied to the country-product network~\cite{caldarelli2012network, mealy2019interpreting}. 
    The algorithm was later shown to be mathematically equivalent to a spectral clustering method that partitions countries and products into groups. Subsequently, Tacchella et al.~\cite{tacchella2012new, tacchella2013economic} proposed a different approach with the Economic Fitness and Complexity (EFC) method, introducing a non-linear iterative framework to capture better the interplay between country diversification and product competitiveness~\cite{cristelli2013measuring, cristelli2015heterogeneous}. The concept of fitness is closely linked to the notion of nestedness in bipartite networks~\cite{mariani2024ranking}, and is a reliable predictor of GDP growth~\cite{cristelli2015heterogeneous}. Later refinements of EFC, such as the regularized Non-Homogeneous Economic Fitness and Complexity (NHEFC) method~\cite{servedio2018new}, 
    were introduced specifically to resolve the convergence instabilities observed in the original algorithm~\cite{pugliese2016convergence,wu2016mathematics}. The regularization parameter $\delta$ is crucial in this regard, as it ensures robust convergence and improves applicability to empirical data. Henceforth, we will consistently refer to the non-homogeneous version of the EFC algorithm, unless otherwise specified.
    
    While the Economic Complexity approach has demonstrated strong empirical relevance in understanding economic growth and trade networks, its mathematical foundations remain an active area of research. Several studies have explored connections between economic complexity indices and established network science methods~\cite{caldarelli2012network, mealy2019interpreting}. Sciarra et al.~\cite{sciarra2020reconciling} recast both ECI and EFC within a common spectral framework, highlighting their structural similarities despite algorithmic differences that produce markedly different country rankings~\cite{cristelli2013measuring}. Other work has investigated the relationships between these indices and standard centrality measures~\cite{piccardi2018complexity, costantini2022measuring, servedio2025fitness}, demonstrating how economic complexity algorithms provide insight into the spectral properties of underlying graphs.
    Extending these algorithms to arbitrary network structures, beyond the canonical bipartite formulation, has significantly broadened their applicability and interpretability. Therefore, a deeper theoretical understanding of these approaches is crucial to expanding their potential applications, as already demonstrated in fields such as ecology~\cite{lin2018nestedness, calo2025measuring, mougi2022predator}.

    Building upon these developments, recent studies have proposed a connection between the EFC algorithm and optimal transport theory~\cite{mazzilli2024equivalence}, particularly in relation to the Sinkhorn--Knopp algorithm~\cite{sinkhorn1967concerning}, which is used to compute doubly stochastic matrices. In this context, economic complexity indices can be interpreted as solutions to optimization problems analogous to those found in optimal transport~\cite{villani2008optimal, cuturi2013sinkhorn}, where cost functions play a central role in defining the equilibrium distributions~\cite{servedio2024economic}. Notably, the iterative structure of the EFC algorithm resembles the scaling procedures used in transport problems, and its associated cost function is associated with Kantorovich potentials~\cite{villani2009optimal}, which characterize the optimal allocation of resources over networks. More generally, the interpretation of economic complexity algorithms as optimization problems~\cite{hartmann2002optimization, arora2015optimization, wright2006numerical} provides a unified framework that clarifies their mathematical structure, reveals properties of their solutions, and facilitates connections with other domains.
    
    This study provides a rigorous optimization-theoretic foundation for both ECI and EFC algorithms. We demonstrate that each minimizes a distinct cost function: ECI is associated with an energy function derived from the network's Laplacian matrix, while EFC is associated with a strictly convex function, for which we provide explicit characterization. We establish global convergence to the fixed point for regular graphs and prove local convergence with explicit rates for general undirected, unweighted graphs.
    Based on this analysis, we propose a gradient-descent modification that significantly accelerates numerical convergence, with potential applications to the broader class of Sinkhorn-Knopp matrix renormalization algorithms.
    Finally, we complement the theoretical analysis with an empirical study of the energetic landscape on a real-world trade network, showing how the two cost functions distribute energy across links and revealing structurally critical regions of the network. This link-level perspective provides both descriptive insight and practical guidance for assessing the organization and fragility of empirical economic systems.

    The results presented in this paper enhance the interpretability of economic complexity algorithms, paving the way for methodological improvements and new applications in network-based data analysis. 
    The reformulation of these indices as cost-minimization problems contributes to a more rigorous theoretical understanding of economic complexity and opens new perspectives for their application beyond traditional bipartite economic networks.


\section{Materials and Methods}

\subsection{Data}
\noindent
For the analysis presented in this study, we primarily rely on synthetic regular graphs with a variable number of nodes $N$ and fixed degree $K$, identical for all nodes. These graphs are artificially generated to match the structural properties required for the analytical investigations developed in the following sections. Since many of our theoretical results are derived under the assumption of regularity, these synthetic graphs serve as a controlled setting to validate and illustrate the analytical findings.

We also consider the well-known Zachary Karate Club network to assess the algorithm’s behavior on real-world data. This dataset, originally collected by Wayne Zachary in 1977~\cite{konect2017zachary,zachary1977information}, represents the social interactions among members of a university karate club. Each node corresponds to a club member, while edges represent observed social ties between them. The network is undirected and is commonly used as a benchmark for community detection algorithms, particularly for identifying the two factions that formed due to a conflict between the club’s instructors.

For the empirical application, we use international trade data from the United Nations COMTRADE database \footnote{\url{https://comtradeplus.un.org}}, classified according to the Standard International Trade Classification (SITC) Revision~2 \cite{united1961standard}. We extract export flows for the year 2015 and construct the country--product bipartite matrix using the standard RCA threshold \cite{RCA}. SITC~Rev.~2 aggregates products into roughly 200 categories, providing a stable and widely used representation of countries' export. This dataset offers a heterogeneous and realistic benchmark for assessing the energetic landscape of ECI and Fitness on real-world bipartite networks.

\subsection{Previous work}
\noindent
In our previous work~\cite{servedio2025fitness, servedio2024economic}, we described two methods for quantifying economic complexity: the Economic Complexity Index (ECI) and the Economic Fitness and Complexity (EFC). Both approaches traditionally operate on the country–product binary matrix $\mathbf{M}$. In this matrix, the entry $M_{cp}$ is set to 1 if country $c$ significantly exports product $p$, and 0 otherwise, as determined by the Revealed Comparative Advantage (RCA) criterion~\cite{RCA}.

For ECI, one defines the diversification of a country and the ubiquity of a product as 
\[
k_c=\sum_{p=1}^{N_p}M_{cp}\quad \text{and} \quad k_p=\sum_{c=1}^{N_c}M_{cp},
\]
and updates the scores iteratively via the method of reflections~\cite{hidalgo2009building}:
\[
F^{(n)}_c=\frac{1}{k_c}\sum_{p=1}^{N_p}M_{cp}Q^{(n-1)}_p,\quad Q^{(n)}_p=\frac{1}{k_p}\sum_{c=1}^{N_c}M_{cp}F^{(n-1)}_c,
\]
with initial conditions $F^{(0)}_c=k_c$ and $Q^{(0)}_p=k_p$. 

We can recast ECI in a spectral framework by writing the iterative process in matrix form using the diagonal matrices $\mathbf{D}$ and $\mathbf{U}$ (with elements $k_c$ and $k_p$, respectively), so that
\[
\vec{F}^{(n)}=\mathbf{D}^{-1}\mathbf{M}\,\vec{Q}^{(n-1)}~~ \text{and}~~ \vec{Q}^{(n)}=\mathbf{U}^{-1}\mathbf{M}^{T}\,\vec{F}^{(n-1)}.
\]
After two iterations, these expressions lead to the transition matrices $\mathbf{N}_1=\mathbf{D}^{-1}\mathbf{M}\mathbf{U}^{-1}\mathbf{M}^{T}$ and $\mathbf{N}_2=\mathbf{U}^{-1}\mathbf{M}^{T}\mathbf{D}^{-1}\mathbf{M}$ whose second eigenvectors (beyond the trivial constant principal eigenvector) capture the nontrivial complexity rankings~\cite{caldarelli2012network,mealy2019interpreting}. 

In contrast, EFC employs a non‐linear map~\cite{tacchella2012new}. 
To ensure robust convergence, we used the regularized Non-Homogeneous version of EFC (NHEFC) given by~\cite{servedio2018new}
\begin{equation}
    \begin{cases}
        {{F}}^{(n)}_c=\delta+\sum_{p=1}^{N_p}M_{cp}/{{S}}^{(n-1)}_{p}\\
        {{S}}^{(n)}_p=\delta+\sum_{c=1}^{N_c}M_{cp}/{{F}}^{(n-1)}_{c}
    \end{cases}
    \label{eq:symmetricNHEFC}
\end{equation}
where $F^{(0)}_c=S^{(0)}_p=1$ and $\delta=10^{-3}$. 

By introducing the adjacency matrix $\mathbf{A}$ and the diagonal matrix $\mathbf{K}$ with $k_c$ and $k_p$ on its diagonal, we can provide a mono-partite representation that recasts both ECI and EFC as iterative maps over a unified vector $\vec{V}$ of dimension $N=N_c+N_p$. 

In particular, we write ECI as
\begin{equation}
    \vec{V}^{(n)} = \mathbf{K}^{-1} \mathbf{A}\; \vec{V}^{(n-1)} ~~\mbox{with}~~~ \vec{V}^{(0)}=\vec{k}.
    \label{eq:eci_transition}
\end{equation}
We similarly write EFC as
\begin{equation}
    \vec{V}^{(n)} = \delta\cdot\vec{1} + \mathbf{A}\cdot (\vec{V}^{(n-1)})^{-1} ~~\mbox{with}~~~ \vec{V}^{(0)} = \vec{1},
    \label{eq:compact_nhefc}
\end{equation}
or explicitly
\begin{equation}
    V^{(n)}_i = \delta + \sum_{j=1}^N \frac{A_{ij}}{V^{(n-1)}_j}~~\mbox{with}~~~ V_i^{(0)} = 1,
    \label{eq:explicit_nhefc}
\end{equation}
thereby generalizing the approach, originally designed for bipartite structures, to any simple-graph structure.
In Eq.~(\ref{eq:compact_nhefc}), the notation $(\vec{V})^{-1}$ stands for taking the reciprocal of each element of the vector $\vec{V}$.
This mono-partite formulation is a key generalization, allowing the ECI and EFC concepts to be applied to any arbitrary network structure, not just the bipartite country-product networks for which they were originally designed~\cite{servedio2024economic}.

\subsection{Cost Functions}
\noindent
In this section, we show that both ECI and EFC can be understood as optimization algorithms, as they can be reformulated as problems of minimizing specific cost functions. 
In many optimization settings, the cost (or objective) function is defined from the outset, and iterative procedures are devised to find its minimum. In our case, the process is reversed: starting from the iterative maps in \Eq{\ref{eq:eci_transition}} and \Eq{\ref{eq:compact_nhefc}}, we reconstruct the corresponding cost functions by integration. 
The cost function associated with ECI relates directly to the standard eigenvector problem and can be interpreted within that framework. For EFC, the connection to optimization has been previously noted due to structural similarities between the fitness algorithm and the Sinkhorn--Knopp algorithm from optimal transport theory~\cite{knight2008sinkhorn}. However, while prior work has identified this formal analogy~\cite{mazzilli2024equivalence}, our study provides the first explicit derivation of EFC's cost function. This explicit formulation enables us to prove the function's convexity rigorously and subsequently develop a more efficient gradient-based algorithm.
In the following, we derive the explicit expression of the cost function associated with the EFC algorithm. Full details of the derivations for both ECI and EFC are provided in Appendix~\ref{app:cost_derivation}.

\subsubsection{ECI}
\noindent
By integrating \Eq{\ref{eq:eci_transition}}, we obtain that the solution of the algorithm is equivalent to finding the minimum of the following cost function:
\begin{equation}
    U(\vec{V}) = - \dfrac{1}{2} \sum_{i, j} A_{ij} V_i V_j + \dfrac{1}{2}\sum_i k_i V_i^2.
    \label{eq:eci_cost_general}
\end{equation}
This function has the same form as the cost function used in standard eigenvector-finding algorithms~\cite{watkins1993some}. 
It is also formally related to a classical harmonic potential~\cite{helms2009potential} or spin-system Hamiltonian, where the first term represents a standard quadratic interaction, and the second acts as a regularization.

We can also express the cost function \Eq{\ref{eq:eci_cost_general}} in terms of the Laplacian matrix $L_{ij} = k_i \delta_{ij} - A_{ij}$ associated with the adjacency matrix $A_{ij}$:
\begin{equation}
    U(\vec{V}) = \frac{1}{4}\sum_{i,j} A_{ij} (V_i - V_j)^2 = \dfrac{1}{2} \sum_{i, j} L_{ij} V_i V_j.
    \label{eq:eci_cost_general_L}
\end{equation}
The minimum of this function occurs when $V_i = V_j$ for all $i$ and $j$, which corresponds to the constant vector. This is also the first eigenvector of the Laplacian matrix, representing the solution that the ECI algorithm approaches after many iterations. This cost function has a natural physical interpretation: it corresponds to the potential energy of a networked system of one-dimensional springs, where nodes represent masses and edges represent springs connecting them.

\subsubsection{EFC}
\noindent
In this case, integrating \Eq{\ref{eq:explicit_nhefc}} gives the following cost function:
\begin{equation}
    U(\vec{V}) = \dfrac{1}{2} \sum_{i, j} A_{ij} \dfrac{1}{V_i V_j} - \sum_i \log\dfrac{1}{V_i} + \delta \sum_i \dfrac{1}{V_i}.
    \label{eq:efc_cost_general}
\end{equation}
The cost function \Eq{\ref{eq:eci_cost_general}} associated with ECI resembles a standard harmonic potential, where the quadratic term promotes positive interactions (i.e., attraction) between connected nodes. In contrast, the cost function of EFC features interaction terms that involve the inverse of the variables. This results in a locally repulsive dynamics, which promotes differentiation between neighboring nodes, opposite to the attractive mechanisms typical of harmonic potentials~\cite{helms2009potential}. The expression \Eq{\ref{eq:efc_cost_general}} also exhibits formal analogies with Kantorovich potentials in optimal transport theory~\cite{villani2009optimal}.

The repulsive nature of the EFC formulation is already evident in the map \Eq{\ref{eq:explicit_nhefc}}, where each fitness value is computed as a weighted average of the inverse of its neighbors. This structure inherently promotes diversity in node values. As a result, and as thoroughly analyzed in~\cite{servedio2025fitness}, fitness centrality is particularly effective in identifying structural vulnerability in networks: it tends to assign high values to nodes connected to weak or fragile ones.

In addition to the interaction term, the cost function includes two regularization terms, both essential for preventing divergence. The first is a logarithmic barrier, which regularizes the contribution of the inverse variables.  This term also reveals that the natural scale of the fitness variables is logarithmic, as already pointed out in several cases \cite{tacchella2012new, mazzilli2024equivalence}. However, this contribution alone is insufficient to prevent the divergence caused by the hyperbolic nature of the interaction. The second regularization term, introduced via the constant $\delta$, ensures full algorithm convergence. The parameter $\delta$ can be interpreted as a Lagrange multiplier that enforces the constraint $\sum_i V_i^{-1} = \text{const}$.

\section{Results}
\label{sec:results}
\noindent
In this section, we analyze the convergence properties of ECI and EFC. For ECI, properties such as the convexity of the potential and the existence of a fixed point follow directly from the fact that ECI is equivalent to computing the second eigenvector of the matrix $\mathbf{K}^{-1} \mathbf{A}$. The following part focuses on the convergence of EFC across different network structures, starting from a simple two-node undirected graph and extending to regular graphs and arbitrary topologies. The analysis establishes local convergence criteria near fixed points and explores the influence of the non-homogeneous parameter $\delta$ on the algorithm's behavior. Finally, we introduce an alternative formulation of the fitness algorithm that follows the gradient of the potential function. This conservative formulation defines the most direct path to the fixed point and significantly reduces the number of iterations required for convergence.

\subsection{Convergence properties of ECI}
\noindent
As previously noted, computing ECI values is equivalent to extracting the eigenvectors of the matrix $\mathbf{K}^{-1} \mathbf{A}$, as defined by the map in \Eq{\ref{eq:eci_transition}}. Since $\mathbf{K}^{-1} \mathbf{A}$ is row-normalized, it acts as a transition matrix. By the Perron–Frobenius theorem, this map converges as $n \to \infty$ to the principal eigenvector of the transition matrix. This eigenvector corresponds to the eigenvalue $\lambda_1 = 1$ and is the constant vector $\vec{V} = (1, \dots, 1)$. This result also follows directly from the cost function in \Eq{\ref{eq:eci_cost_general_L}}, which attains its global minimum when all $V_i$ are equal. At this point, the cost is $U(\vec{V}) = 0$. Moreover, since the cost function in Eq.~(\ref{eq:eci_cost_general_L}) is a quadratic form defined by the Laplacian network matrix, it is convex over the entire domain. 
Therefore, the map \Eq{\ref{eq:eci_transition}} has a unique fixed point, and the Perron–Frobenius theorem ensures convergence. 

Since this fixed point (the constant vector) does not provide a meaningful solution of the problem of ranking countries or products, the original formulation of ECI~\cite{hidalgo2009building} prescribes stopping the iteration after a fixed number of steps. As discussed in several works~\cite{sciarra2020reconciling, servedio2024economic, hidalgo2021economic}, the intermediate steps of the iteration retain information about the subdominant eigenvectors, particularly the second one. As a result, ECI can be obtained in two equivalent ways: (i) by iterating the map \Eq{\ref{eq:eci_transition}} for a fixed number of steps, or (ii) by directly computing the second eigenvector of the transition matrix $\mathbf{K}^{-1} \mathbf{A}$. This eigenvector maximizes the partitioning of the network, in line with the interpretation of ECI as a spectral algorithm for community detection~\cite{sciarra2020reconciling}. 
This observation clarifies the geometric nature of the algorithm: ECI minimizes the cut of the network, tending to group countries with similar export baskets together. This clustering behavior explains why ECI is mathematically distinct from, and orthogonal to, measures intended to capture diversification capabilities~\cite{sciarra2020reconciling}.

\subsection{Convergence properties of EFC}

\subsubsection{Uniqueness of solution via convexity analysis}
\label{sec:uniqueness}

\begin{figure*}
    \centering
    \includegraphics[width=1\linewidth]{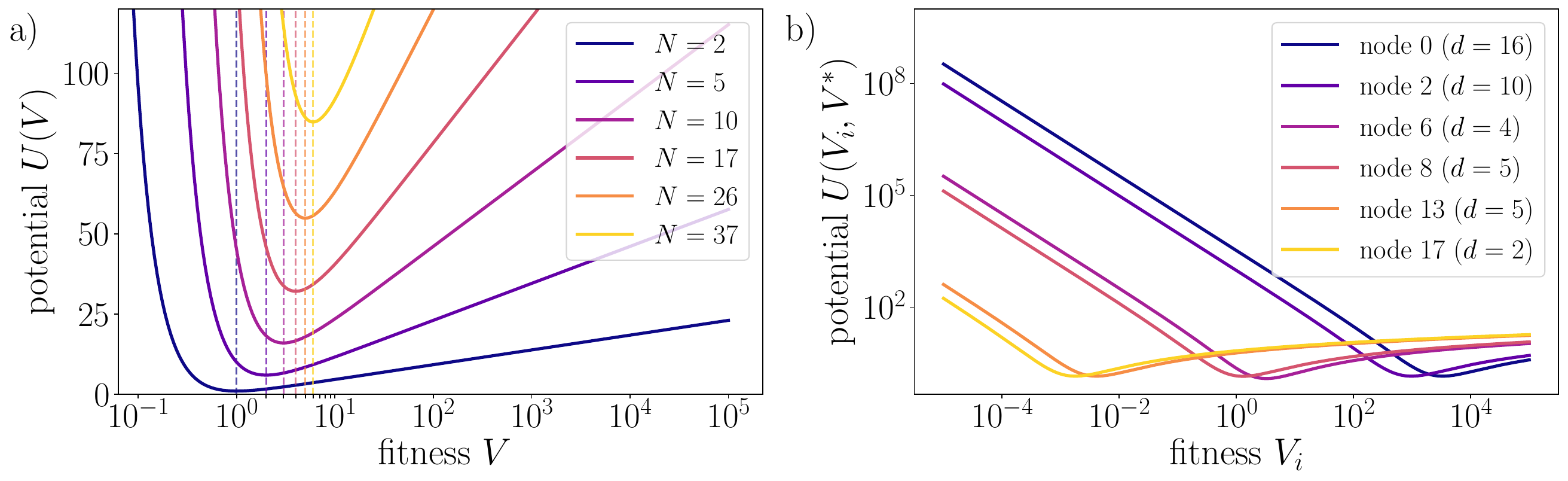}%

    \caption{\textbf{Energetic landscape of the fitness algorithm.} 
    \textbf{a)} Cost function from Eq.~(\ref{eq:efc_cost_general}) for regular complete graphs with varying number of nodes $N$ and $K = N-1$. The potential $U(V)$ is plotted as a function of a single variable $V$, since by symmetry all node variables can be set equal. The curves for different values of $N$ demonstrate the convexity of the potential. Each curve exhibits a global minimum corresponding to the fixed point in Eq.~(\ref{eq:fixed_point_regular}), which for small $\delta$ can be approximated as $V^* \approx \sqrt{K}$. 
    \textbf{b)} Cost function around the fixed point $V^*$ for the ZKC network~\cite{zachary1977information}, plotted as a function of the fitness value $V_i$ for specific nodes. Each curve shows how the potential varies around the fixed point $V_i^*$ for that node. Nodes with higher fitness display larger values of $V_i^*$, i.e., their minima lie farther to the right. The legend reports the node degrees $d$ for comparison. As shown, the highest fitness nodes contribute most to the total potential, while low-fitness nodes provide only local corrections. Each curve exhibits a unique, global minimum. The curves are convex, though the logarithmic scale on the vertical axis may create a misleading appearance of concavity in some regions.
    }
    \label{fig:fig1}
\end{figure*}

\noindent
To establish the uniqueness of the solution to the EFC equations, we investigate the properties of the potential function $U(\vec{V})$ defined in \Eq{\ref{eq:efc_cost_general}}. We consider the form of the potential $U(\vec{Z})$ expressed in terms of the natural fitness variables $Z_i = -\log V_i$, as derived in Appendix~\ref{app:cost_derivation}:
\begin{equation}
U(\vec{Z}) = \dfrac{1}{2} \sum_{i, j} A_{ij} e^{Z_i} e^{Z_j} - \sum_i Z_i + \delta \sum_i e^{Z_i}
\label{eq:NHEFC_cost_Z}
\end{equation}
Specifically, we demonstrate its strict convexity by analyzing its Hessian matrix. The second partial derivatives of $U$ with respect to the variables $Z_i$ yield the Hessian matrix elements:
\begin{equation}
    H_{ij} = \frac{\partial^2 U}{\partial Z_i \partial Z_j} =
        \delta_{ij} e^{Z_i} \left(\sum_{k} A_{ik} e^{Z_k} + \delta\right) + A_{ij} e^{Z_i}e^{Z_j},
        \label{eq:hessian_explained}
\end{equation}
where $\delta_{ij}$ represents the Kronecker delta.

This Hessian matrix allows for a useful interpretation rooted in graph theory. Let us consider an associated weighted graph where the edge weights are given by $W_{ij} = A_{ij} e^{Z_i} e^{Z_j}$ (non-zero only if $A_{ij}=1$) and the strength of each node $i$ is $S_i = \sum_k W_{ik} = e^{Z_i} \sum_k A_{ik} e^{Z_k}$. With these definitions, the Hessian matrix $\mathbf{H}$ (with components $H_{ij}$) can be decomposed as:
$$\mathbf{H} = \mathbf{Q} + \mathbf{\Delta}.$$
Here, $\mathbf{Q}$ is the signless Laplacian matrix of the weighted graph, defined by $Q_{ij} = \delta_{ij} S_i + W_{ij}$. From a well-established property in spectral graph theory, the signless Laplacian $\mathbf{Q}$ of any simple graph (weighted or unweighted) is positive semi-definite; that is, all of its eigenvalues are non-negative ($\lambda(\mathbf{Q}) \ge 0$). The second term, $\mathbf{\Delta}$, is a diagonal matrix with entries $\Delta_{ii} = \delta e^{Z_i}$ and zeros elsewhere ($\Delta_{ij} = \delta_{ij} \delta e^{Z_i}$).

Since the EFC algorithm assumes $\delta > 0$ and the exponential function ensures $e^{Z_i} > 0$ for all $i$, the diagonal entries of $\mathbf{\Delta}$ are strictly positive. Adding a diagonal matrix with strictly positive entries ($\mathbf{\Delta}$) to a positive semi-definite matrix ($\mathbf{Q}$) results in a matrix ($\mathbf{H}$) that is positive definite. This means all eigenvalues of the Hessian matrix $\mathbf{H}$ are strictly positive ($\lambda(\mathbf{H}) > 0$) for any state $\vec{Z}$.

A fundamental result from multivariate calculus states that if the Hessian matrix of a function is positive definite throughout its domain, the function is strictly convex. Therefore, the potential function $U(\vec{Z})$ is strictly convex. A strictly convex function possesses, at most, one point where its gradient vanishes. This stationary point corresponds precisely to the condition $\dot{Z}_i = 0$ in the dynamical system formulation (\Eq{\ref{eq:efc_field_z}}), which in turn defines the fixed point of the EFC iterative map \Eq{\ref{eq:efc_field}}. Consequently, the strict convexity of $U(\vec{Z})$ guarantees that this fixed point is not only a minimum but the unique global minimum of the potential function, ensuring the uniqueness of the EFC solution.

The role of \(\delta\) can also be analyzed through its influence on the cost function bounds. 
From \Eq{\ref{eq:explicit_nhefc}}, using \( k_{\max} \) as the maximum node degree in the network, we obtain the constraint \( \delta \leq V_i \leq \delta + k_{\max}/\delta \). 
In terms of the logarithmic variables \( Z_i \), this translates to the bounds \( -\log(\delta + k_{\max}/\delta) \leq Z_i \leq -\log(\delta) \).  
By incorporating these inequalities into \Eq{\ref{eq:EFCpotential}} and retaining only the leading terms, we establish the global bounds on the potential function,  
\begin{equation}
    -N\log(1/\delta) \leq U \leq E/\delta^2,
    \label{eq:potential_boundary}
\end{equation}  
where \( N \) is the number of nodes and \( E \) is the number of links in the network. 
This result demonstrates that \(\delta\) effectively regulates the potential function, preventing divergence and ensuring that \( U \) remains bounded.
From an optimization perspective, this explains why the non-homogeneous map does not suffer from the pathological convergence issues known to affect the original fitness algorithm. The regularization ensures the potential is strictly convex, guaranteeing a unique stable solution even for non-bipartite or non-nested topologies where the original map would fail.

Figure~\ref{fig:fig1} illustrates the shape of the potential function $U(\vec{Z})$ for the EFC algorithm. The left panel shows the case of regular graphs of different sizes, where all nodes are symmetric and share the same fitness value (Eq.~(\ref{eq:fixed_point_regular})). In this case, the potential is plotted as a function of a single variable, confirming its strict convexity.
The right panel shows the potential landscape for the Zachary Karate Club network. Here, each node has a distinct fitness value. We visualize the potential $U(\vec{Z})$ level curves by varying one variable at a time around the fixed point. These cross-sections show that the potential is convex along all directions in the fitness space. Moreover, the shape of each curve reveals how much each node contributes to the total potential. Nodes with high fitness (indicated by the location of the minimum in their respective potential sections) induce the steepest variations in the potential. This suggests that central or structurally important nodes contribute more significantly to the system's energetic configuration, while peripheral nodes have a more negligible effect and contribute only local corrections.

\subsubsection{Solution in the case of two connected nodes}
\label{sec:N=2}

\begin{figure*}
    \centering
    \includegraphics[width=1\linewidth]{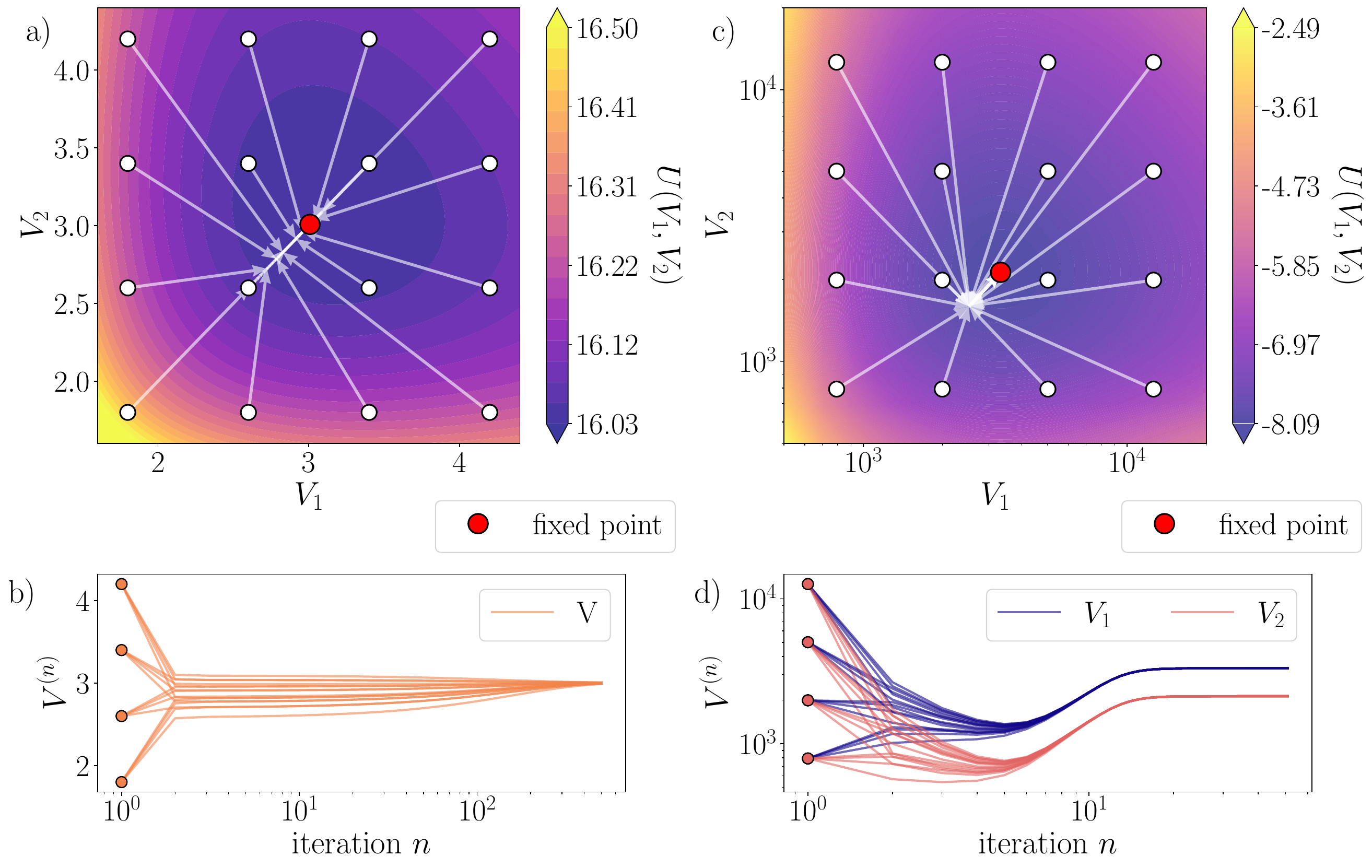}%

    \caption{\textbf{Convergence of trajectories in the phase space.} 
    \textbf{a)} Bidimensional phase space of variables $V_1$ and $V_2$ for a complete graph with $N = 10$. The colormap shows the value of the potential, with the global minimum corresponding to the fixed point $(V_1^*, V_2^*) \approx (3, 3)$ (Eq.(\ref{eq:fixed_point_regular})). White points indicate several initial conditions, while the lines show the convergence of the original map defined in Eq.~(\ref{eq:explicit_nhefc}). Trajectories are shown every ten steps. 
    \textbf{b)} Convergence of several trajectories, starting from different initial conditions, toward the fixed point. Only one variable, $V$, is plotted, since by the symmetry of the graph, all node variables are equivalent.  Trajectories are shown every two steps, due to the oscillatory behavior of the map.
    \textbf{c)} Bidimensional phase space of $V_1$ and $V_2$ for the ZKC network~\cite{zachary1977information}. The global minimum of the potential corresponds to the fixed point $(V_1^*, V_2^*) \approx (3.2, 2.2) \times 10^3$. White points indicate various initial conditions, and the lines show the trajectories under Eq.~(\ref{eq:explicit_nhefc}), plotted every two steps. 
    \textbf{d)} Convergence of several trajectories to the fixed point for variables $V_1$ and $V_2$, starting from different initial conditions.
    }
    \label{fig:fig2}
\end{figure*}

\noindent
The specific case of a network with only two connected nodes ($N=2$) offers a tractable scenario where the algorithm's convergence can be analyzed in full detail (refer to Appendix~\ref{app:N=2} for a complete treatment). By examining the dynamics over two consecutive steps, it is found that each component $V_j$ evolves independently according to the same one-dimensional iterative map: $V^{(n+2)}_j = f(V^{(n)}_j)$, where the function is defined as $f(x) = \delta + x/(1+\delta x)$. For any positive value of the parameter $\delta$, this function $f(x)$ has derivative $0 < f'(x) < 1$ for $x>0$, guaranteeing the existence of a unique, globally attracting fixed point within the positive domain. This fixed point is given by:
\[
x^* = \frac{1}{2} \left( \sqrt{4+\delta^2} + \delta \right).
\]
Consequently, for $\delta > 0$, the sequence of vectors $\vec{V}^{(n)}$ is guaranteed to converge to the unique equilibrium state $\vec{V}^* = (x^*, x^*)$ regardless of the positive starting point $\vec{V}^{(0)}$. This fixed point $\vec{V}^*$ also corresponds to the unique global minimum of the associated potential function $U(\vec{V})$. The analysis highlights the critical stabilizing role of $\delta$: if $\delta$ is set to zero, the function becomes $f(x) = x$, and convergence is lost; the system typically exhibits persistent period-2 oscillations instead of settling at the fixed point $(1,1)$. Thus, $\delta > 0$ acts like a damping mechanism, ensuring asymptotic stability rather than mere stability.

\subsubsection{Solution in the case of a regular graph}
\label{sec:regular_graph}

\begin{figure*}
    \centering
    \includegraphics[width=\linewidth]{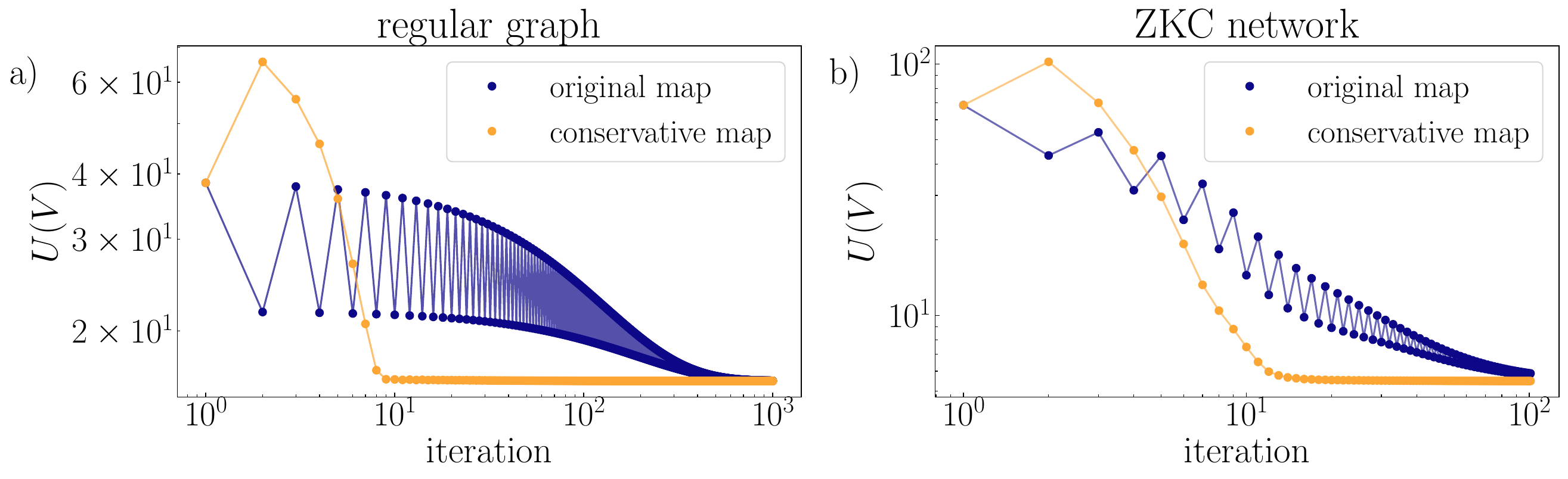}%

    \caption{\textbf{Efficient convergence of the conservative map compared to the original map.} 
    \textbf{a)} Trajectories of the potential $U(\vec{V})$ (Eq.~(\ref{eq:efc_cost_general})) for a complete graph with $N = 10$. Following the original map, the dynamics exhibit strong period-2 oscillations, indicating algorithm instability. In contrast, the trajectory computed by descending the cost function gradient using the conservative map (Eq.~(\ref{eq:efc_map_z})) shows no oscillatory behavior and converges more efficiently to the fixed point. It requires significantly fewer steps to reach a stable solution. 
    \textbf{b)} Trajectories of the potential for the ZKC network. Also in this case, the conservative map converges considerably faster and avoids oscillations. Due to the formal analogy between EFC and the Sinkhorn--Knopp algorithm~\cite{sinkhorn1967concerning}, the conservative map offers a more efficient alternative for matrix renormalization.
    } 
    \label{fig:fig3}
\end{figure*}

\noindent
In the specific case where the underlying network corresponds to a regular graph, meaning every node $i$ has the same degree $K = \sum_j A_{ij}$, the convergence properties of the iterative map can be analytically determined (more details are provided in Appendix~\ref{app:regular}). The analysis relies on tracking the minimum, $V^{(n)}_m$, and maximum, $V^{(n)}_M$, values of the fitness vector components at each iteration $n$. By establishing recursive inequalities, it is shown that the evolution of these bounds is governed by an auxiliary one-dimensional map $X^{(n+1)} = f_K(X^{(n)})$. Provided that the parameter $\delta > 0$, this map possesses a unique, globally attracting fixed point. This structure rigorously proves that for any positive initial condition $\vec{V}^{(0)}$, the iterative process
$$V^{(n)}_i = \delta + \sum_j \frac{A_{ij}}{V^{(n-1)}_j}$$
converges exponentially fast to a unique fixed point $\vec{V}^*$, where all components are identical and equal to
\begin{equation}
    V^* = \sqrt{K+\tfrac{1}{4}\delta^2} + \delta.
    \label{eq:fixed_point_regular}
\end{equation}
This fixed point $\vec{V}^*$ represents the unique positive solution to the stationarity condition $V_i = \delta + \sum_j A_{ij}/V_j$ and also corresponds to the unique global minimum of the associated cost function $U(\vec{V})$ within the positive orthant. The positivity of $\delta$ is crucial, as the argument for convergence and uniqueness breaks down when $\delta=0$. 

The specific shape of the cost function for the EFC algorithm in the case of a regular graph is shown in Fig.~\ref{fig:fig1}a. The potential curves highlight the fixed point, corresponding to the function's global minimum. These curves also clearly confirm the strict convexity of the cost function across the entire fitness space. Additionally, Fig.~\ref{fig:fig2}a displays the potential landscape in the two-dimensional phase space defined by the first two components of the fitness vector, $V_1$ and $V_2$, for a regular graph with $N = 10$ and degree $K = 9$. The figure shows the trajectories of the algorithm starting from various initial conditions, all of which converge to the fixed point, located at $V \approx \sqrt{K} = 3$.

\subsubsection{Local and general convergence for a general graph}
\label{sec:general_graph}

\noindent
For general graph structures, while global convergence is not guaranteed by the methods used for regular graphs, the local stability of the algorithm's unique fixed point $\vec{V}^*$ can be established using Lyapunov stability theory (more details in Appendix~\ref{app:general_graph}). The approach utilizes the potential function $U(\vec{V})$, whose unique minimum coincides with $\vec{V}^*$, as a Lyapunov function candidate. The analysis focuses on the change in the potential over one iteration, $\Delta U = U(\vec{F}(\vec{V})) - U(\vec{V})$, for states $\vec{V}$ confined to a sufficiently small neighborhood around the fixed point $\vec{V}^*$. By performing a Taylor expansion of $\Delta U$ and carefully analyzing its terms, particularly relating the gradient term $\nabla U(\vec{V}) \cdot (\vec{F}(\vec{V})-\vec{V})$ to the step size and bounding the second-order (Hessian) term using properties of the fixed point and inequalities like Cauchy-Schwarz, it can be demonstrated that $\Delta U$ is strictly negative for any $\vec{V} \neq \vec{V}^*$ within this neighborhood, provided $\delta > 0$. To leading order in the displacement $\vec{D} = \vec{F}(\vec{V}) - \vec{V}$, the change is found to be approximately
$$U(\vec{F}(\vec{V})) - U(\vec{V}) \approx - \frac{\delta}{2} \sum_i \frac{D_i^2}{(V_i^*)^3}$$
Since this expression is strictly negative for any non-zero displacement $\vec{D}$ (given $\delta > 0$), it confirms that $U(\vec{V})$ strictly decreases along the iterative trajectory as it approaches $\vec{V}^*$. This rigorously establishes the local asymptotic stability of the unique fixed point $\vec{V}^*$ for the dynamics on general graphs when $\delta > 0$.

While a general proof of convergence is difficult for arbitrary network structures, this property can be investigated empirically by analyzing the behavior of multiple trajectories initialized from different starting points. We carried out this analysis for the Zachary Karate Club network, as shown in Fig.~\ref{fig:fig2}b. The results indicate that, regardless of the initial condition in the phase space, all trajectories converge to the same fixed point. Although a formal proof is lacking, this evidence suggests that global convergence may hold more generally. At the same time, we cannot rule out the existence of non-stationary orbits (e.g., periodic ones) that attract the trajectories of \Eq{\ref{eq:explicit_nhefc}} for suitable choices of the initial configuration.

\subsubsection{Reformulation of the map via gradient descent}

\begin{figure*}
    \centering
    \includegraphics[width=\linewidth]{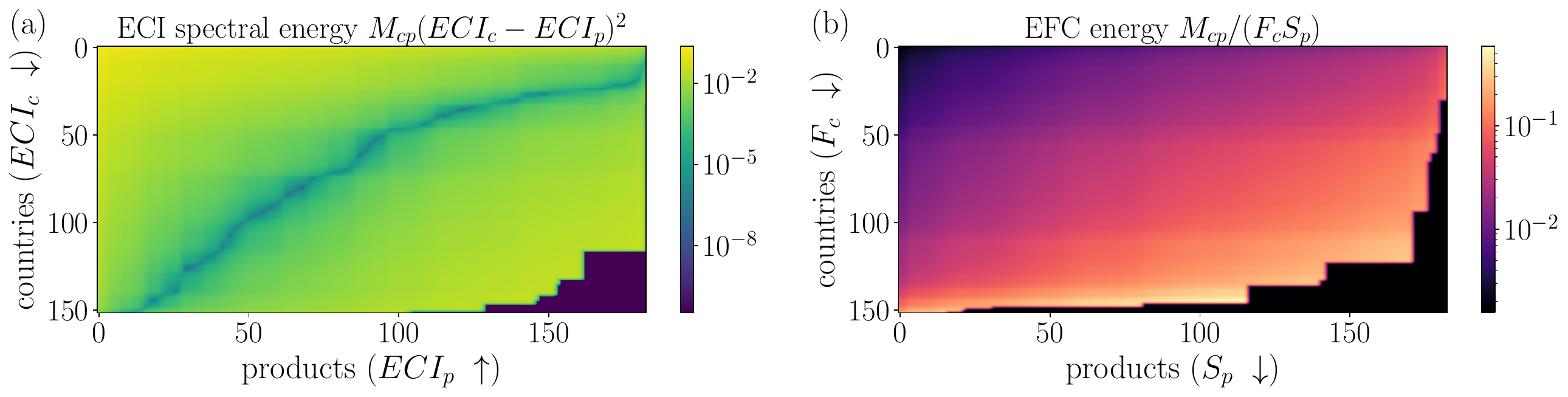}%
    \caption{\textbf{Energetic landscape of ECI and EFC on the trade network.} Link-wise energy for the country--product matrix (UN COMTRADE, SITC Rev.~2, 2015), with rows and columns sorted by the corresponding scores. For visualization purposes, energy values are smoothly interpolated within the non-zero ``frontier'' of the matrix, i.e., the region to the left of the boundary beyond which all entries are zero. \textbf{(a)} ECI energy, $E_{cp}\propto M_{cp}(ECI_c-ECI_p)^2$, is lowest along the diagonal, consistent with its Laplacian structure, and increases for links between nodes with distant spectral positions. \textbf{(b)} EFC energy, $E_{cp} \propto M_{cp}/(F_c S_p)$, concentrates along the ``frontier'' of the matrix, highlighting structurally critical links involving low-fitness countries or low-simplicity products. ECI thus reflects smooth spectral variation, while EFC identifies points of fragility more relevant for targeted perturbations.}
    \label{fig:fig4}
\end{figure*}

\noindent
The reformulation of the fitness algorithm as the minimization of a cost function has practical computational advantages, as shown in Appendix~\ref{app:cost_derivation}. When the dynamics are expressed in the logarithmic variables $Z_i = -\log V_i$, the associated vector field becomes irrotational. This allows us to define a conservative (i.e., gradient-driven) map that follows the steepest descent of the potential function $U(\vec{Z})$, leading to faster convergence. The update rule is:
\begin{equation}
Z_i^{(n)} = Z_i^{(n-1)} + 1 - \delta e^{Z_i^{(n-1)}} - e^{Z_i^{(n-1)}}\sum_{j=1}^N A_{ij} e^{Z_j^{(n-1)}}
\label{eq:efc_map_z}
\end{equation}
This iteration differs from the standard map in Eq.~(\ref{eq:explicit_nhefc}), as it explicitly follows the cost function gradient in the transformed variables. In Fig.~\ref{fig:fig3}, we compare the convergence behavior of both maps. The conservative map exhibits significantly faster convergence and eliminates the oscillatory behavior observed in the original formulation. For practical computations, this gradient-based update rule should be preferred. If necessary, the results can be mapped back to the original variables via $V_i = e^{-Z_i}$.

It is worth noting that, by introducing a scaling factor $\eta$ in front of the gradient in Eq.~(\ref{eq:efc_map_z}), the basin of attraction of the fixed point under this modified dynamics expands as $\eta$ decreases, eventually covering the entire phase space in the limit $\eta\to 0$. Consequently, the robustness of the algorithm can be enhanced arbitrarily.

Given the formal equivalence between the EFC algorithm and the Sinkhorn–Knopp algorithm~\cite{mazzilli2024equivalence, sinkhorn1967concerning}, which is used for matrix renormalization and the computation of doubly stochastic matrices, the conservative map we propose can also be seen as an efficient alternative in this broader algorithmic class.

\subsubsection{Application to real trade networks}

\noindent
We finally apply our framework to a real-world bipartite system: the country--product export network from the UN COMTRADE database \footnote{\url{https://comtradeplus.un.org}} (year 2015), classified according to SITC Rev.~2~\cite{united1961standard}. After thresholding export flows using the standard RCA criterion~\cite{RCA}, we obtain a binary matrix $M_{cp}$ with 152 countries and 183 products. We compute the scores of both algorithms---$(ECI_c,ECI_p)$ for ECI and $(F_c,S_p)$ for EFC---and evaluate the energetic contribution of each observed link: $E_{cp}\propto M_{cp}(ECI_c-ECI_p)^2$ for ECI (Eq.~\ref{eq:eci_cost_general_L}) and $E_{cp} \propto M_{cp}/(F_c S_p)$ for EFC (Eq.~\ref{eq:efc_cost_general}). Figure~\ref{fig:fig4} displays these values as heatmaps aligned with the structure of the trade matrix.

Panel (a) shows that the ECI energy landscape reproduces the geometry of a Laplacian quadratic form: the lowest-energy links lie close to the diagonal, where countries and products have similar ECI values, while high-energy contributions appear far from it. This structure reflects the spectral nature of the ECI cost function, which penalizes connections between nodes occupying distant positions in the eigenvector-based embedding.

The EFC landscape in panel (b) exhibits a different and more heterogeneous structure. High-energy links concentrate along the ``frontier’’ of the matrix, typically involving low-fitness countries or low-simplicity products. This is consistent with the nonlinear amplification mechanism of the EFC algorithm: rare connections that tie highly capable countries to complex products, or peripheral countries to their few essential exports, contribute disproportionately to the potential. These links are structurally critical, since their removal would rapidly isolate low-fitness countries or disconnect the most complex products from the network~\cite{servedio2025fitness}.

This energetic perspective naturally suggests attack or pruning strategies based on link energy. Removing the highest-energy links under the ECI potential would mainly affect off-diagonal connections, but these links are typically redundant for highly diversified countries or sparsely populated for low-ECI products, resulting in limited structural impact. In contrast, removing high-energy links according to the EFC potential would directly target the structural weak points of the network: the essential exports of low-fitness countries and the few connections sustaining highly complex products. Such a strategy would more rapidly destabilize the system, potentially eliminating entire nodes. Hence, the Fitness energy identifies connections whose removal is expected to produce the most pronounced effects on the resilience of the trade network.

\section{Discussion}
\label{sec:discussion}
\noindent
This study establishes the mathematical foundations of two economic complexity measures, {ECI} and {EFC}, by reformulating them as optimization problems. This approach unifies these algorithms with well-established network science techniques and provides a deeper theoretical understanding and significant practical improvements. Our findings demonstrate that the iterative algorithms for both {ECI} and {EFC} can be derived as procedures to minimize specific cost functions, revealing the underlying mechanics of each framework.

A key result of our analysis is the contrasting nature of the cost functions associated with {ECI} and {EFC}. The {ECI} cost function is a quadratic form related to the network's Laplacian matrix, analogous to a classical harmonic potential. This formulation promotes attraction, where connected nodes tend to have similar values, which aligns with the established interpretation of ECI as a spectral algorithm for network partitioning. In stark contrast, the {EFC} cost function features interaction terms involving the inverse of the node variables, resulting in a locally repulsive dynamic. This repulsive mechanism promotes differentiation among neighboring nodes. This explains the effectiveness of fitness centrality in identifying structural vulnerabilities, as it tends to assign high fitness values to nodes connected to weaker ones. More broadly, it establishes a rigorous connection with well-known optimization problems and contributes to understanding the relationship between Economic Fitness and Complexity and the broader framework of optimal transport theory.

Furthermore, our derivation of the {EFC} cost function clarifies the roles of its components. The function reveals that the natural scale of the fitness variables is logarithmic, a property observed empirically but not formally derived. Crucially, the analysis highlights the role of the regularization parameter $\delta$. This parameter acts as a damping mechanism that ensures the algorithm's convergence by preventing the divergence caused by the hyperbolic nature of the interactions. Theoretically, the most significant contribution is the proof of the uniqueness of the {EFC} solution. By transforming the potential into its natural logarithmic variables and analyzing its Hessian matrix, we demonstrated that the potential is strictly convex. This strict convexity guarantees a unique global minimum, which corresponds to the algorithm's fixed point, thus resolving any ambiguity about the uniqueness of the solution. 

The optimization framework also offers significant practical and methodological advancements. For {EFC}, we have shown that the dynamics expressed in logarithmic variables are irrotational. This implies that the most efficient computational path to the solution is a direct gradient descent of the potential function. This leads to a new ``conservative map'' (Eq.~(\ref{eq:efc_map_z}) in the original text) that converges significantly faster and avoids the sawtooth oscillations characteristic of the original map. 
Moreover, given the formal connection between the EFC algorithm and the Sinkhorn-Knopp algorithm, used for matrix renormalization and the computation of doubly stochastic matrices, the conservative map introduced here may also provide an effective computational strategy within that broader class of problems. Future work should explore whether this gradient-based formulation can systematically improve the convergence speed, stability, and theoretical understanding of Sinkhorn-type algorithms, particularly in the context of optimal transport and large-scale network renormalization.

Interpreting economic complexity algorithms through potential functions opens the way to a novel perspective on their application. 
In economic networks, the potential function provides an interpretable energetic landscape, capturing both local and global structural forces. As illustrated in Fig.~\ref{fig:fig4}, the analysis of a real-world trade network shows how the energetic framework translates directly into empirical insight. The distribution of link-wise energy highlights structurally critical regions of the export matrix, revealing which connections sustain the organization and resilience of the system. By identifying links that contribute significantly to the potential, this approach offers a quantitative tool for diagnosing fragility and understanding how local disruptions may propagate through the network.

Viewed from a broader perspective, these empirical patterns also motivate a more general interpretation of the potential function as a device for assessing systemic stability and the network’s response to shocks. Given the demonstrated strict convexity of the potential, the energetic landscape is characterized by a single, unique global minimum, representing the system's sole optimal configuration. At any other point, the potential $U(\vec{V})$ quantifies the ``stress'' or deviation from this ideal state. This picture re-frames economic resilience not as the depth of a potential well, but as the steepness of the landscape around the minimum; a steeper potential implies stronger restorative forces that pull the system back to its optimum after a shock. Consequently, structural transformation is reconceptualized not as a jump between different stable states but as a dynamic change of the landscape itself, where an alteration in the network's topology reshapes the cost function and shifts the location of the unique minimum. Finally, the geometry of this landscape can inform measures of systemic risk, as a ``flatter'' potential in certain directions would indicate a system with weak restorative forces, vulnerable to large and inefficient deviations from its optimal configuration.
This representation could provide powerful insights into the organization and resilience of the economic fabric, offering a new tool for empirical investigations in regional development, economic geography, and beyond.

Finally, this work broadens the applicability and the theoretical interpretation of economic complexity measures. By reformulating the underlying algorithms as optimization problems, we establish formal connections to broader scientific frameworks, including Kantorovich potentials in optimal transport theory and spectral methods in graph theory through the Laplacian matrix. Crucially, expressing the dynamics in a mono-partite setting removes the constraint of bipartite trade networks, allowing these measures to be applied to arbitrary graph structures. This generalization opens the way for their adoption in diverse contexts, such as ecological systems, infrastructure networks, and spatial economic analysis, significantly extending their scope and relevance.

\vspace{2mm}
\noindent
\textbf{Author contributions:}
V.D.P.S. conceptualized and led the study, including methodology, validation, and initial draft writing. A.B. contributed to methods and software development and created all visualizations. P.B. contributed to the formal analysis. All authors collaborated on the formal analysis and the final manuscript.
All authors have read and agreed to the published version of the manuscript.

\vspace{2mm}
\noindent
\textbf{Funding:}
This work was co-financed by the Austrian Federal Ministry for Innovation, Mobility and Infrastructure (BMIMI) as part of the project GZ 2021-0.664.668 and the Oesterreichische Nationalbank as part of their funding programme for Austrian economic research.

\vspace{2mm}
\noindent
\textbf{Conflicts of interest:}
The authors declare no conflict of interest.

\vspace{2mm}
\noindent
\textbf{Abbreviations:}
The following abbreviations are used in this manuscript:\\
\begin{tabular}{@{}ll}
    ECI & Economic Complexity Index\\
    EFC & Economic Fitness and Complexity\\
    NHEFC & Non-Homogeneous Economic Fitness and Complexity\\
    RCA & Revealed Comparative Advantage
\end{tabular}
\vspace{2mm}
\begin{acknowledgments}
\noindent We acknowledge valuable discussions with E.~Cal\`o, G.~De~Marzo, D.~Mazzilli, A.~Patelli, and A.~Tacchella.
\end{acknowledgments}

\appendix

\section{Derivation of Cost Functions}
\label{app:cost_derivation}
\noindent
Here, we provide the analytical details of the computations for both ECI and EFC algorithms. We start with the simplest case, namely ECI.

\subsection{ECI}
\noindent
To express the cost function, we must formulate \Eq{\ref{eq:eci_transition}} as a map. This involves subtracting $V_i^{(n-1)}$ from both sides of the equations and considering the left-hand side $V_i^{(n)} - V_i^{(n-1)} \approx \frac{d}{dn}V_i = \dot{V_i}^{(n-1)}$ as an approximate derivative. The resulting map is expressed as:
\begin{equation}
    \dot{V_i}^{(n-1)} = \dfrac{1}{k_i} \sum_{j=1}^N A_{ij} V_j^{(n-1)} - V_i^{(n-1)}
    \label{eq:eci_field}
\end{equation}
Here, the dot represents the derivative with respect to the iteration index $^{(n)}$, and $N = N_c + N_p$ denotes the total number of nodes in the network, irrespective of their specific nature. The right-hand side of the equation can be interpreted as the vector field driving the dynamics to convergence.

It is convenient to introduce a change of variable to make the network's adjacency matrix symmetric. 
This is required to compute the cost function of the map. 
Introducing the variables $Z_i = \sqrt{k_i} V_i$, \Eq{\ref{eq:eci_field}} reads as:
\begin{equation}
\dot{Z}_i = \sum_{j=1}^N \dfrac{A_{ij}}{\sqrt{k_i k_j}} Z_j - Z_i
\label{eq:eci_field_2}
\end{equation}
The potential can now be obtained by integrating this map, but the Schwartz condition must be satisfied. 
This is equivalent to requiring that the vector field is irrotational (to be more precise, the 2-form derived from the vector field 1-form is zero):
\begin{equation}
\dfrac{\partial \dot{Z_i}}{\partial Z_h} = \dfrac{\partial \dot{Z_h}}{\partial Z_i}
\label{eq:schwarz}
\end{equation}
Given that the full adjacency matrix $A_{ij}$ is symmetric by construction, both quantities are equal to $\dfrac{A_{ih}}{\sqrt{k_i k_h}} + \delta_{ih}$. 

This ensures the existence of a potential function in the variables $Z_i$ with the property that $- \nabla_{Z_i} U = \dot{Z_i}$. 
From this we recover the explicit expression by integrating \Eq{\ref{eq:eci_field_2}}:
\begin{equation}
U(\vec{Z}) = - \dfrac{1}{2} \sum_{i, j} \dfrac{A_{ij}}{\sqrt{k_i k_j}} Z_i Z_j + \dfrac{1}{2}\sum_i Z_i^2.
\label{eq:ECI_potential_Z}
\end{equation}
The factor $1/2$ avoids counting each link twice since the matrix $A_{ij}$ is symmetric. 
Replacing back the original variables $V_i = Z_i / \sqrt{k_i}$, the cost functions is expressed as:
\begin{equation}
U(\vec{V}) = - \dfrac{1}{2} \sum_{i, j} A_{ij} V_i V_j + \dfrac{1}{2}\sum_i k_i V_i^2
\label{eq:ECI_potential_V}
\end{equation}
We remark that the interpretation as a potential function is possible only using the natural variables $Z_i$. 
The minimum of \Eq{\ref{eq:ECI_potential_Z}} coincides with the minimum of \Eq{\ref{eq:ECI_potential_V}}, but the dynamics leading to the minimization may not be equivalent.

\subsection{EFC}
\noindent
For EFC, after removing the isolated nodes the map equivalent to \Eq{\ref{eq:eci_field}} is: 
\begin{equation}
\dot{V_i}^{(n-1)} = \delta + \sum_{j=1}^N \frac{A_{ij}}{V^{(n-1)}_j} - V_i^{(n-1)}
\label{eq:efc_field}
\end{equation}
The vector field associated with the system is not irrotational, as \Eq{\ref{eq:schwarz}} are not satisfied. Through an appropriate change of variable $Z_i = - \log V_i$, \Eq{\ref{eq:efc_field}} can be transformed as
\begin{equation}
\dot{Z_i} = 1 - \delta e^{Z_i} - e^{Z_i}\sum_{j=1}^N A_{ij} e^{Z_j}
\label{eq:efc_field_z}
\end{equation}
This map is now irrotational, and the change of variable naturally reveals the intrinsic logarithmic scale of EFC. 
Integrating this expression gives:
\begin{equation}
U(\vec{Z}) = \dfrac{1}{2} \sum_{i, j} A_{ij} e^{Z_i} e^{Z_j} - \sum_i Z_i + \delta \sum_i e^{Z_i}
\label{eq:EFCpotential}
\end{equation}
Finally, \Eq{\ref{eq:efc_cost_general}} can be obtained by replacing the original variables $V_i = e^{-Z_i}$.

\section{EFC: Solution in the case N=2}
\label{app:N=2}
\noindent
In the case of a simple network consisting of \( N=2 \) connected nodes, the equations defining the iterative map can be directly formulated. Specifically, we have  
\[
\begin{cases}
V^{(n+1)}_1 = \delta + \frac{1}{V^{(n)}_2}, \\ 
V^{(n+1)}_2 = \delta + \frac{1}{V^{(n)}_1}
\end{cases}
\]
along with the associated function  
\[
U(\vec{V}) = \frac{1}{V_1V_2} + \log V_1 + \log V_2 + \frac{\delta}{V_1} + \frac{\delta}{V_2}.
\]
From these expressions, it follows that  
\[
V^{(n+2)}_j = f(V^{(n)}_j), \quad j=1,2, \quad \text{where } f(x) = \delta + \frac{x}{1+\delta x}.
\]
The iterative dynamics given by \( x_{n+1} = f(x_n) \) can be analyzed using a Lamerey diagram. Observing that the function \( f(x) \) has a unique fixed point in the interval \( [0,+\infty) \), given by  
\[
x^* = \frac{1}{2} \left( \sqrt{4+\delta^2} + \delta \right),
\]
and that its derivative satisfies  
\[
0 < f'(x) < 1, \quad f([0,+\infty)) = [\delta, 1/\delta),
\]
it follows that the sequence \( x_n \) converges monotonically to \( x^* \): if \( x_0 \in [0,x^*) \), then \( x_n \) is increasing and approaches \( x^* \), while if \( x_0 > x^* \), the sequence is decreasing and converges to \( x^* \).  

Furthermore, the function \( U(\vec{V}) \) reaches a stationary point within the domain \( [0,+\infty)^2 \) if and only if \( \vec{V} = (x^*, x^*) \), as determined by the condition \( \nabla U(\vec{V}) = 0 \). Since \( U(\vec{V}) \to +\infty \) when \( \vec{V} \) approaches the boundary of \( [0,+\infty)^2 \), this unique stationary point is necessarily the absolute minimizer of \( U \). Consequently, both sequences \( U(\vec{V}^{(2n)}) \) and \( U(\vec{V}^{(2n+1)}) \) converge to \( \min U \), ensuring that the entire sequence \( U(\vec{V}^{(n)}) \) also converges to its minimum value.  

It is important to emphasize the critical role of the positive parameter \( \delta \) in ensuring convergence toward the fixed point. In particular, when \( \delta = 0 \), there remains a unique fixed point, given by \( \vec{V}^* = (1,1) \). However, in this case, since the map effectively alternates between two steps at a time and reduces to the identity function, \( f(x) = x \), the system does not exhibit true convergence. Instead, the phase space becomes entirely populated by periodic orbits, each with a period of two, oscillating around \( \vec{V}^* \).  

From a dynamical perspective, the parameter \( \delta \) plays a role analogous to that of linear friction in a mechanical system. Just as friction causes the minimum of a potential function to transition from being merely stable to asymptotically stable, the presence of \( \delta > 0 \) ensures that trajectories are progressively damped and eventually settle at the fixed point, eliminating the periodic behavior observed in the frictionless case.  

\section{EFC: Regular graphs}
\label{app:regular}
\noindent
The iterative scheme is defined by
\[
V^{(n)}_i = \delta + \sum_j \frac{A_{ij}}{V^{(n-1)}_j},
\]
where the matrix elements \(A_{ij}\) are such that, for some integer \(K\in \{1, \ldots, N-1\}\),
\[
K_i := \sum_j A_{ij} = K \qquad \forall\, i.
\]
In particular, if \(K=N-1\), the graph is complete (clique). Notice that, independently of the initial condition \(\vec{V}^{(0)} \in (0,+\infty)^N\), the iterates are uniformly bounded. One easily verifies that
\begin{equation}
\label{Vn<}
\delta \le V^{(n)}_i \le \delta + \frac{K}{\delta} \qquad \forall\, n \ge 1 \quad \forall\, i.
\end{equation}

To analyze the convergence properties of the map, we define the minimum and maximum values at iteration \(n\) as
\[
V^{(n)}_m := \min_j V^{(n)}_j, \qquad V^{(n)}_M := \max_j V^{(n)}_j.
\]
Then, by the structure of the update, it follows that
\[
\delta + \frac{K}{V^{(n)}_M} \le V^{(n+1)}_i \le \delta + \frac{K}{V^{(n)}_m} \qquad \forall\, i.
\]
This immediately implies the bounds
\[
\delta + \frac{K}{V^{(n)}_M} \le V^{(n+1)}_m \quad \text{and} \quad V^{(n+1)}_M \le \delta + \frac{K}{V^{(n)}_m}.
\]
By iterating these inequalities once more, we obtain
\begin{equation}
\label{VmM}
V^{(n+2)}_m \ge f_K(V^{(n)}_m), \qquad V^{(n+2)}_M \le f_K(V^{(n)}_M),
\end{equation}
where we have introduced the function
\[
f_K(x) = \delta + \frac{Kx}{K+\delta x}.
\]
It is instructive to note that
\[
0 < f_K'(x) < 1, \quad f_K([0,+\infty)) = [\delta, K/\delta),
\]
which implies that the iterative process
\begin{equation}
\label{Xn=}
X^{(n+1)} = f_K(X^{(n)})
\end{equation}
behaves in a manner analogous to the two-node case. In particular, for any initial condition \(X^{(0)}>0\), the sequence \(X^{(n)}\) converges to the unique fixed point \(V^*\) of \(f_K\) in \([0,\infty)\), where
\[
V^* = \sqrt{K+\tfrac{1}{4}\delta^2} + \delta.
\]

To further formalize this convergence, denote by \(X^{(n)}_m\) and \(Y^{(n)}_m\) the orbits starting from initial values \(X^{(0)}_m \le V^{(0)}_m\) and \(Y^{(0)}_m \le V^{(1)}_m\), respectively. Then, by the inequalities in \eqref{VmM} and the monotonicity of \(f_K\),
\[
V^{(2n)}_m \ge X^{(n)}_m, \qquad V^{(2n+1)}_m \ge Y^{(n)}_m \qquad \forall\, n\ge 0.
\]
These bounds can be rigorously established via induction. For instance, assuming that \( V^{(2n)}_m \ge X^{(n)}_m\), we have
\[
V^{(2n+2)}_m \ge f_K(V^{(2n)}_m) \ge f_K(X^{(n)}_m) = X^{(n+1)}_m.
\]
A similar argument, using orbits \(X^{(n)}_M\) and \(Y^{(n)}_M\) starting from \(X^{(0)}_M \ge V^{(0)}_M\) and \(Y^{(0)}_M \ge V^{(1)}_M\), leads to
\[
V^{(2n)}_M \le X^{(n)}_M, \qquad V^{(2n+1)}_M \le Y^{(n)}_M \qquad \forall\, n\ge 0.
\]
In summary, the sequences \(\{V^{(n)}_m\}\) and \(\{V^{(n)}_M\}\) are bounded respectively from below and above by trajectories of the dynamical system defined in \eqref{Xn=}. Consequently, we obtain
\begin{equation}
\label{VnV*}
\lim_{n\to \infty} \vec{V}^{(n)} = \vec{V}^* \qquad \forall\, \vec{V}^{(0)} \in (0,+\infty)^N,
\end{equation}
with \(\vec{V}^* := (V^*,\ldots,V^*)\).

It is worthwhile to note that, from \eqref{VmM} and the lower bound in \eqref{Vn<}, one deduces that
\[
\begin{split}
&V^{(n)}_M - V^{(n)}_m \le \frac{K^2\left(V^{(n-2)}_M - V^{(n-2)}_m\right)}{(\delta V^{(n-2)}_M +K) (\delta V^{(n-2)}_m +K)}\\
&\le \frac{K^2}{(K+\delta^2)^2} \left(V^{(n-2)}_M - V^{(n-2)}_m\right).
\end{split}
\]
Since \(\frac{K^2}{(K+\delta^2)^2} < 1\), this inequality shows that the convergence in \eqref{VnV*} is exponentially fast in time. 
Moreover, applying \eqref{VnV*} to the stationary case, we deduce that the fixed point equation
\begin{equation}
\label{fixpocl}
V_i =  \delta + \sum_j \frac{A_{ij}}{V_j}
\end{equation}
admits a unique solution \(\vec{V} = \vec{V}^*\) in \((0,+\infty)^N\).

As discussed at the end of Subsection \ref{sec:N=2} in the explicit case \(N=2\), the presence of the positive parameter \(\delta\) is crucial for ensuring convergence to a unique fixed point. When \(N>2\) and \(\delta=0\), the argument leading to \eqref{VnV*} fails because the function \(f_K(x)\) becomes the identity \(f_K(x) = x\). Moreover, unlike the special case \(N=2\), it is not even clear whether \(\vec{V}^* = (\sqrt{K},\ldots,\sqrt{K})\) is the unique fixed point.

Regarding the relation to the cost function, we observe that
\[
\frac{\partial U}{\partial V_i} = - \frac{1}{V_i^2}  \sum_j \frac{A_{ij}}{V_j} + \frac{1}{V_i} -\frac{\delta}{V_i^2} =
\]
\[
= - \frac{1}{V_i^2} \left( \sum_j \frac{A_{ij}}{V_j} - V_i + \delta \right).
\]
Hence, the condition for a stationary point, \(\nabla U(\vec{V}) = 0\), is equivalent to the fixed point equation \eqref{fixpocl}. Since \(U(\vec{V})\) diverges to \(+\infty\) as \(\vec{V}\) approaches the boundary of \((0,+\infty)^N\), we conclude that the fixed point \(\vec{V}^*\) is indeed the unique absolute minimizer of \(U\) in \((0,+\infty)^N\).

\section{EFC: local convergence for general graphs}
\label{app:general_graph}
\noindent
We write the dynamics in vector form as
\[
\vec{V}^{(n+1)} = \vec{F}(\vec{V}^{(n)})
\]
where $\vec{F}(\vec{V})$ denotes the vector-valued function with components
\[
\vec{F}_i(\vec{V}) = \delta + \sum_j \frac{A_{ij}}{V_j}
\]

Drawing upon the results established in Section \ref{sec:uniqueness}, the mapping $\vec{F}$ possesses precisely one fixed point, which coincides with the unique minimizer $\vec{V}^*$ of the cost function $U$. We aim to demonstrate that $W=U-\min U$ serves as a Lyapunov function for the equilibrium $\vec{V}^*$. According to standard stability theory, this would imply that the equilibrium is asymptotically stable.

Evidently, there exists a ball $B = B_\varepsilon(\vec{V}^*)$ of radius $\varepsilon$ centered at $\vec{V}^*$ within which $W\geq 0$ and $W(\vec{V}) = 0$ if and only if $\vec{V} = \vec{V}^*$. To complete the proof, we must show that, for sufficiently small $\varepsilon$,
\[
U(F(\vec{V})) - U(\vec{V}) < 0 \qquad \forall\, \vec{V}\in B \setminus\{\vec{V}^*\}
\]
(which implies that the same property holds with $W$ in place of $U$). To this end, we observe that
\[
F_i(\vec{V}) = V_i - V_i^2 \frac{\partial U}{\partial V_i} (\vec{V}), 
\]

Therefore, defining $\vec{D} = \vec{F}(\vec{V}) - \vec{V}$, we find 
\[
\nabla U (\vec{V}) \cdot \vec{D} = - \sum_i \frac{D_i^2}{V_i^2} 
\]

By Taylor's formula,
\[
\begin{split}
&U(F(\vec{V})) - U(\vec{V}) \\ 
&= \nabla U (\vec{V}) \cdot \vec{D} + \int_0^1\!\rmd\lambda \, \lambda \vec{D} \cdot D^2 U(\vec{V} + \lambda \vec{D})\vec{D} \\ 
&= - \sum_i \frac{D_i^2}{(V_i^*)^2}  + \frac 12 \vec{D} \cdot D^2 U(\vec{V}^*)\vec{D} + \mathcal{O}(\varepsilon)|\vec{D}|^2,\\
&\forall\, \vec{V}\in B,
\end{split}
\]
where in the last equality, we have utilized the fact that if $\vec{V}\in B$, then $|\vec{V} + \lambda \vec{D} - \vec{V}^*| \leq C \varepsilon$, since $\vec{F}$ is regular and $\vec{V}^*$ is a fixed point. Furthermore,
\[
(D^2U)_{ij}(\vec{V}) = \frac{A_{ij}}{V_i^2V_j^2} + \left[\frac{2}{V_i^3} \left(\sum_s \frac{A_{ij}}{V_s} + \delta\right) - \frac{1}{V_i^2}\right] \delta_{ij} 
\]

The Hessian evaluated at the fixed point simplifies to
\[
(D^2U)_{ij}(\vec{V}^*) = \frac{A_{ij}}{(V_i^*)^2(V_j^*)^2} + \frac{\delta_{ij} }{(V_i^*)^2}
\]
(note that it is positive definite, by the same argument used for \Eq{\ref{eq:hessian_explained}}), yielding
\[
\frac 12 \vec{D} \cdot D^2 U(\vec{V}^*)\vec{D} = \frac 12 \sum_{i,j} \frac{A_{ij}D_iD_j}{(V_i^*)^2(V_j^*)^2} + \frac 12 \sum_i \frac{D_i^2}{(V_i^*)^2}    
\]

Applying the Cauchy–Schwarz inequality and leveraging the fixed point property of $\vec{V}^*$, we obtain 
\[
\begin{split}
&\frac 12 \sum_{i,j} \frac{A_{ij}D_iD_j}{(V_i^*)^2(V_j^*)^2} = \frac 12 \sum_{i,j} \frac{A_{ij}}{V_i^*V_j^*} \frac{D_i}{V_i^*} \frac{D_j}{V_j^*}\\
&\leq \frac 14 \sum_{i,j} \frac{A_{ij}}{V_i^*V_j^*} \left( \frac{D_i^2}{(V_i^*)^2} + \frac{D_j^2}{(V_j^*)^2} \right) \\
&= \frac 12 \sum_{i,j} \frac{A_{ij}}{V_i^*V_j^*} \frac{D_i^2}{(V_i^*)^2} = \frac 12 \sum_i \frac{D_i^2}{(V_i^*)^3}  \sum_j \frac{A_{ij}}{V_j^*}\\
&= \frac 12 \sum_i \frac{D_i^2}{(V_i^*)^3} (V_i^* - \delta) = \frac 12 \sum_i \frac{D_i^2}{(V_i^*)^2} - \frac\delta 2 \sum_i \frac{D_i^2}{(V_i^*)^3}
\end{split}
\]

In conclusion, denoting $\displaystyle V_M^* = \max_iV_i^*$, we establish
\[
\begin{split}
U(F(\vec{V})) - U(\vec{V}) &\leq - \frac\delta 2 \sum_i \frac{D_i^2}{(V_i^*)^3} + \mathcal{O}(\varepsilon)|\vec{D}|^2 \\&\leq \left(-\frac{\delta}{2(V_M^*)^3} + \mathcal{O}(\varepsilon)\right) |\vec{D}|^2,\\
&\forall\, \vec{V}\in B
\end{split}
\]
which implies $U(F(\vec{V})) - U(\vec{V})<0$ provided $\varepsilon$ is sufficiently small and $\vec{D} \neq \vec{0}$ (i.e., $\vec{V}\neq \vec{V}^*$).

\end{document}